\documentclass[reprint,superscriptaddress,showpacs,aps,twocolumn]{revtex4-1}
\usepackage{amsmath}
\usepackage{graphicx}
\usepackage{dcolumn}% Align table columns on decimal point
\usepackage{bm}% bold math
\usepackage{epsfig}
\usepackage{color}
\usepackage{tikz}
\usepackage{subcaption}
\usepackage{comment}
\usepackage[normalem]{ulem}
\usepackage[colorlinks=true,urlcolor=blue,linkcolor=blue,
            citecolor=blue]{hyperref}
\newcommand \beq{\begin{eqnarray}}
\newcommand \eeq{\end{eqnarray}}
\newcommand \bea{\begin{eqnarray}}
\newcommand \eea{\end{eqnarray}}

\def\simge{\mathrel{%
       \rlap{\raise 0.511ex \hbox{$>$}}{\lower 0.511ex \hbox{$\sim$}}}}
\def\simle{\mathrel{
       \rlap{\raise 0.511ex \hbox{$<$}}{\lower 0.511ex \hbox{$\sim$}}}}

\begin{document}
\title{Electronic energy gap closure and metal-insulator transition in dense liquid hydrogen.}
\author{Vitaly Gorelov}
\affiliation{Maison de la Simulation, CEA, CNRS, Univ. Paris-Sud, UVSQ, Universit{\'e} Paris-Saclay, 91191 Gif-sur-Yvette, France}
\author{David M. Ceperley}
\affiliation{Department of Physics, University of Illinois, Urbana, Illinois 61801, USA}
\author{Markus Holzmann} 
\affiliation{Univ. Grenoble Alpes, CNRS, LPMMC, 3800 Grenoble, France}
\affiliation{Institut Laue Langevin, BP 156, F-38042 Grenoble Cedex 9, France}
\author{Carlo Pierleoni}
\affiliation{Maison de la Simulation, CEA, CNRS, Univ. Paris-Sud, UVSQ, Universit{\'e} Paris-Saclay, 91191 Gif-sur-Yvette, France}
\affiliation{Department of Physical and Chemical Sciences, University of L'Aquila, Via Vetoio 10, I-67010 L'Aquila, Italy}

\date{\today}

\begin{abstract}
Using Quantum Monte Carlo (QMC) calculations, we 
investigate the insulator-metal transition observed in
liquid hydrogen at high pressure.
Below the critical temperature of the transition from the molecular
to the atomic liquid, 
the fundamental electronic gap closure occurs abruptly, with a small
discontinuity reflecting the weak first-order transition in the 
thermodynamic equation of state.
Above the critical temperature, 
molecular dissociation sets in while the gap is still open. When the gap closes, the
decay of the off-diagonal reduced density matrix shows that the liquid
enters a gapless, but localized phase: there is a cross-over between the insulating
and the metallic liquids.
Compared to different DFT functionals, 
our QMC calculations 
provide larger values for the fundamental gap and the electronic density of states
close to the band edges, 
indicating that optical properties from DFT potentially benefit from
error cancellations.
%\CP{should we mentioned that this is related to the observed agreement with experiments?}
\end{abstract}

%\pacs{}

\maketitle

\section{Introduction}

The insulator-metal (IM) transition in liquid hydrogen has been an outstanding issue in high pressure physics. Initially, the first order transition from an insulating molecular to a metallic monoatomic fluid, called the plasma-phase transition (PPT), was predicted theoretically to occur below a critical temperature based on chemical models \cite{Landau1943,Norman1970,Ebeling1985,Saumon1989}. %\vitaly{Besides, there was proposed that above some critical temperature dielectric liquid should continuously transform to a metal} \cite{Landau1943,Ebeling1985}.  
Experimentally high P-T conditions necessary to observe what is now called the liquid-liquid phase transition (LLPT)  can be achieved in two ways: using dynamic or static compression. Dynamically, hydrogen can be compressed with shock waves, following the time-varying changes in pressure, the metallic states can be detected via electrical, optical and density measurements \cite{Nellis1992,Weir1996,Nellis1999,Celliers2000,Loubeyre2004,Loubeyre2012,Fortov2007,Knudson2015,Celliers2018}. Metallic liquid hydrogen can also be investigated in diamond anvil cell (DAC), using controlled laser heating at constant volume \cite{Dzyabura2013,Ohta2015,Zaghoo2016,Zaghoo2017,McWilliams2016,Jiang2020}. 
A rapid change in the reflectivity has been observed with both techniques, but
inconsistencies between different experimental results remain.

Overall, most experiments conclude that metallization of liquid hydrogen occurs in two steps: entering first into the absorbing semiconductor regime followed by the rapid increase of reflectivity and the IM transition. However, the behaviour of the fundamental gap remains uncertain: whether there is a Mott-like temperature activated transition accompanied by a continuous band overlap, or gap closure is discontinuous and coincides with the LLPT. 
Here we investigate this question using Quantum Monte Carlo (QMC)
calculations of the fundamental
gap in the vicinity of the transition.

Experimentally, the most direct information on the IM transition can be achieved via the conductivity measurements. The first experimental work that directly determined conductivity using resistance measurements in liquid hydrogen was carried out with shock wave compression in a gas-gun experiment \cite{Nellis1992,Weir1996,Nellis1999}. To achieve high pressures, the initial shock was split into multiple, relatively weak shocks reverberating in hydrogen between two sapphire anvils. The resistance was measured using electrodes at the hydrogen/anvil interface. %, the electrodes, in turn, were connected to the oscilloscope through a battery-charged isolated capacitor. When a shock wave transits the liquid hydrogen between the electrodes, the liquid becomes conducting and the capacitors discharge through hydrogen allowing the resistance measurements. Pressure was determined via the measured mass velocity of the initial shock and the Hugoniot equation of state (EOS) of the sapphire anvil. The other thermodynamic parameters such as density and temperature were determined from different EOS of hydrogen \cite{KERLEY1983,Ross1998} which result in a large temperature uncertainty.
Based on the minimum metallic conductivity of $2000$ $(\Omega cm)^{-1}$ the IM transition was placed at 140 GPa and 2600 K, although temperature was inferred from a model equation of state. To determine the energy gap the authors fit the conductivity in the range $93-120$ GPa with a model appropriate for a liquid semiconductor with the thermally activated conductivity that depends on the mobility gap and the limiting value of conductivity.

Shock compression can as well be a laser-driven process \cite{Celliers2000,Celliers2018}. The setup is almost identical to the one in gas-gun experiment, except that the shock wave is created by laser irradiation of the pusher (Al, Be or Cu) which is transmitted to the liquid hydrogen/deuterium. %The shock velocity and the reflectance are measured with another lasers. 
%Thermodynamic parameters are inferred from the velocity data based on the known EOSs. Different EOSs result again in the large uncertainty on temperature. 
%The latest results on  
The IM transition in liquid deuterium is predicted to be first order with the critical temperature in the range 1100 K$<T_C<$3000 K and a critical pressure about 200 GPa \cite{Celliers2018}. Based on optical measurements, two transition boundaries have been identified: first, the sample becomes opaque, corresponding to the onset of absorption at the energies of the detecting laser $\sim 2eV$, then the reflectivity increases by $30\%$, which is attributed to the IM transition. 
The band gap was estimated using empirical relations for the refractive index of semiconductors. %The refractive index of deuterium in turn is determined based on the known relation as a function of the density in order to match the calculated \vitaly{apparent} velocity of the shock to the observed one. 

In a similar experiment a shock wave in deuterium was created using an electromagnetic current pulse \cite{Knudson2015}. Absorption appeared in the same P-T range as Cellier et al. experiment \cite{Celliers2018}. In this set up the reflectivity decrease upon pressure release was monitored and an abrupt jump was observed between 280 and 305 GPa. The temperature range (inferred from a theoretical EOS) was between 1000 K and 1800 K. The band gap was not measured directly, but based on the energy of absorption onset ($\sim 2.3$ eV) was qualitatively compared to the reanalysed data of Weir et al. \cite{Weir1996} and to first-principles density functional theory (DFT) predictions.

%\vitaly{dont forget all the other papers on dynamic compression, like \cite{Loubeyre2004,Loubeyre2012,Fortov2007}}

Hydrogen is a very diffusive material, therefore, during static compression, it is difficult to achieve the high temperatures required to observe the IM transition. However, using short pulsed-laser heating it was possible to reach up to 3000 K in a DAC with compressed liquid hydrogen \cite{Dzyabura2013,Ohta2015,Zaghoo2016,Zaghoo2017,Zaghoo2018}. By increasing the laser power, a plateau in temperature between 1100-2200 K and 90-160 GPa \cite{Dzyabura2013, Ohta2015,Zacharias2016} accompanied by the increase of reflectivity and decrease in optical transmission \cite{Zaghoo2016} was interpreted as being due to the latent heat, a signature of the first order phase transition. However, a finite element analysis (FEA) of the pulsed-laser heated DAC predicts the latent heat necessary to reach the plateau to be rather large ($\sim 2 eV/atom$), in contrast to the theoretical predictions at the LLPT ($\sim 0.035$ eV) \cite{Morales2010}. The plateaus were alternatively interpreted by other authors as the onset of hydrogen absorption \cite{Knudson2015,Celliers2018,Goncharov2017}. Measured reflectivity reached saturation at higher temperatures than the plateau \cite{Zaghoo2017}. % \vitaly{SHALL WE MENTION THIS? Based on the Drude fit, at saturation, hydrogen is predicted to be largely atomic and degenerate, in contrast to the semiconductor model. However, below the saturation the nature of the liquid is non-free-electron like \cite{Zaghoo2018PRE}. } \CP{I would cut this.}

Using long pulsed-laser heating, another experimental group observed a similar two-stage transition: an anomalous temperature behaviour and the onset of absorption followed by the rapid increase of the reflectivity \cite{McWilliams2016,Jiang2020}. However, the P-T conditions ascribed to this transitions are in disagreement with the previous DAC experiments \cite{Dzyabura2013,Ohta2015,Zaghoo2016,Zaghoo2017,Zaghoo2018}. The authors used Tauc's relation \cite{Tauc1968} to describe the observed absorption profile $\alpha(\omega)$ of the semiconducting liquid hydrogen: $\alpha(\omega) \propto (\hbar \omega - E_g)^2/\hbar \omega$, where $E_g$ is the inferred band gap. %in the later we will asses the validity of this model. 

To model the IM transition in liquid hydrogen several theoretical studies have been made \cite{Pierleoni2016,Knudson2015,Morales2010,Mazzola2018,Lorenzen2010,Holst2011,Morales2013liquid,Norman2018,Gorelov2019,Lu2019,Hinz2020} based on Born-Oppenheimer Molecular Dynamics (BOMD) and Path Integral Molecular Dynamics (PIMD)  %predict the presence of the first-order transition between insulating molecular and conductive monoatomic fluid. 
The XC approximation within DFT strongly influences the pressure and temperature of the transition \cite{Knudson2018,Ramakrishna2020,Geng2019,Lu2019,Hinz2020,Morales2013liquid}. More reliable QMC-based methods (CEIMC and QMC-based molecular dynamics \cite{Pierleoni2016,Mazzola2018}) predict a transition line that is in agreement with the experimental observation of the reflective sample in most of the experiments except the one by Knudson et al. \cite{Knudson2018}. 

The electronic properties necessary to identify the IM transition, such as optical conductivity, reflectivity and absorption can be computed within DFT \cite{Rillo2019,Pierleoni2016,Morales2010,Holst2011,Lorenzen2010,Hinz2020,Lu2019} by the Kubo-Greenwood formula \cite{Kubo1957,Greenwood1958}. Based on the HSE density functional and nuclear trajectory from CEIMC \cite{Rillo2019}, the DC conductivity and reflectivity jump coincides with the dissociation transition, which together with the onset of absorption agrees with most experiments \cite{Celliers2018,Zaghoo2018,McWilliams2016}. However, changing the XC approximation in the optical calculation gives rather different results on optical properties and shifts the IM transition line \cite{Lu2019,Hinz2020}. Therefore, considering a correlated many-body theory, such as QMC, can give an accurate prediction of optical properties and might further serve as a benchmark for single electron theories. 

In the past, using the QMC method and the many-body Kubo formula \cite{Kubo1957,Lin2009}, the electrical conductivity has been computed for liquid hydrogen at temperatures above the critical point and found a good agreement with the experimental results available at the time  \cite{Weir1996,Nellis1999}. However, to address the IM transition it is necessary to have calculations for temperatures below and above the critical point of the LLPT. 
In this paper,  
we perform a fully consistent characterisation of the IM transition in liquid hydrogen extending to liquids our
recently developed method for accurately computing
energy gaps within QMC for ideal \cite{Yang2020} and thermal crystals \cite{Gorelov2020}.

The paper is organized as follow. Section \ref{sec:thermeth} describes the methods used in the present study and section \ref{sec:results} reports our results on the closure of the fundamental gap of liquid hydrogen together with the benchmark of several DFT XC functionals and the discussion of optical properties. Section \ref{sec:conclusions} contains our conclusions.

%Below critical temperature the electronic momentum distribution was shown to be discontinuous based on QMC and was associated with a change of electronic localization from non-Fermi to Fermi liquid character \cite{Pierleoni2018}

\section{Theoretical method}\label{sec:thermeth}

Here we report results of an extensive study of the band gap closure of hydrogen near the LLTP using a recently developed QMC based method \cite{Yang2020,Gorelov2020}. We have studied liquid hydrogen along three isotherms: $T=900$, 1500 and 3000 K. Nuclear quantum effects were addressed using imaginary time path integrals for the protons. All systems considered had $N_p=54$ protons at constant volume
and periodic boundary conditions. Optimized Slater-Jastrow-Backflow trial wave functions 
for the electrons with twist averaged boundary conditions have been used for the CEIMC calculations; details of the CEIMC simulations are reported in Ref. \cite{Pierleoni2016}. Averages over ionic positions for band gaps were obtained using at least 16 statistically independent nuclear configurations from the CEIMC trajectories. 

For a given nuclear configuration we perform several reptation quantum Monte Carlo (RQMC) calculations with a varying number of electrons,
$N_e=N_p+N$ with $N \in [-6,6]$. We use an imaginary-time projection $t=2.0$ Ha$^{-1}$ and time step $\tau=0.01$ Ha$^{-1}$ and a $6\times6\times6$ Monkhorst-Pack grid of twists. Electronic size effects on the gap are treated as discussed in \cite{Yang2020}. %\CP{The gaps of quantum liquid are computed by first averaging the electronic total energies for different number of electrons according to the thermal distribution over the nuclear configuration and then applying the grand-canonical twist-averaged boundary conditions (GCTABC) \cite{Yang2020,Gorelov2020}, I FIND THIS SENTENCE CONFUSING, DO YOU MEAN THAT WE HAVE TO AVERAGE OVER NUCLEAR CONFIGURATION BEFORE TAKING THE ENERGY DERIVATIVE WITH DENSITY? WHAT IS NOT CLEAR TO ME IS THAT YOU SAY GENERICALLY THAT AFTER AVERAGING WE PERFORM THE GC-TABC CONDITIONS: WHAT ARE THESE CONDITIONS?}. In the grand-canonical ensemble the fundamental gap is defined as the difference in chemical potentials 
between adding and removing electrons, $\mu_+$ and $\mu_-$, respectively (see
SM of Ref. \cite{Gorelov2020} for more details),
\beq
\Delta_{gc}=\mu_+ - \mu_-\simeq\left.\frac{d \langle e \rangle_{N_p}}{d n_e}\right|_{N_p^+}-\left.\frac{d \langle e \rangle_{N_p}}{d n_e}\right|_{N_p^-},
\label{eq:gapApp}
\eeq
where $e$ is the energy density, expressed as a function of electronic density $n_e=N_e/V$,
$\langle \cdots \rangle_{N_p}$ denotes the average over the Born-Oppenheimer 
energy surface of the undoped $N_e=N_p$ system,
and the discontinuity in the derivative
is computed at the equilibrium density $n_e=n_p=N_p/V$.

Optical properties were calculated within single electron theory using the linear response Kubo-Greenwood formula \cite{Kubo1957,Greenwood1958}. Thermodynamic averages of optical properties were computed with the HSE XC and Williams-Lax \cite{Williams1951,Lax1952} semiclassical approximations using at least 16 uncorrelated configurations from the CEIMC run. More details on these calculations of optical properties are given in Ref. \cite{Rillo2019}. To achieve a better convergence of the DFT gaps we reanalyzed some of the HSE-DFT calculations reported in Ref. \cite{Rillo2019} with an increased k-point grid ($8 \times 8 \times 8$). 

To correct the band gap error when computing the optical properties within DFT, one can rigidly shift the unoccupied eigenvalues by the QMC-DFT gap difference, $\Delta_{sc} = \Delta_{QMC}-\Delta_{DFT}$. This defines the ``scissor'' correction. Alternatively, it is possible to shift the obtained Kubo-Greenwood conductivity directly by the $\Delta_{sc}$. We verified that the two procedures are, in fact, equivalent.

\section{Results}\label{sec:results}

\subsection{The fundamental gap}

\begin{figure}[t]
\includegraphics[width=\columnwidth]{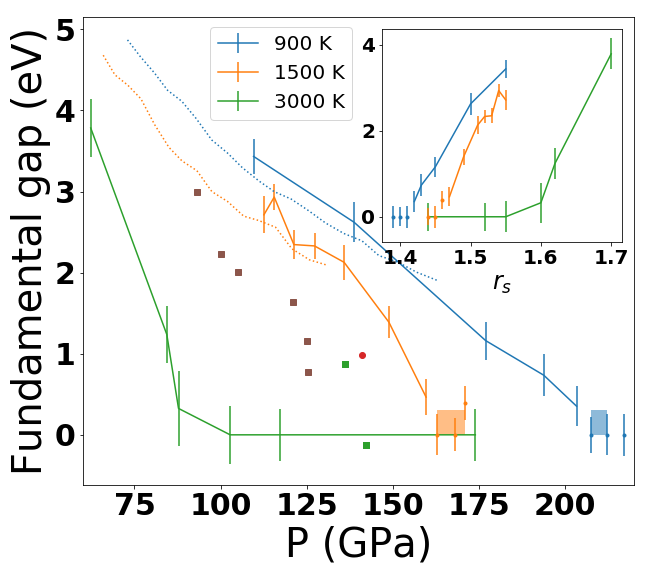}
\caption{\small{The fundamental energy gap of liquid hydrogen along the isotherms: T = 900 K, 1500 K and 3000 K as a function of pressure. Inset: the same gap as a function of r$_s$, a measure of density. The lines connect the gap data only up to the molecular-atomic transition region. The colored rectangles show the coexistence region of the LLPT according to Ref. \cite{Pierleoni2017}. The dotted lines are the gaps reported by Cellier et at. \cite{Celliers2018}. The brown and green squares are the results of Nellis et al. for temperatures 2000-3000 K \cite{Nellis1999} reanalyzed in ref. \cite{Knudson2018}.  The red circle is the gap reported by McWilliams et al. at 2400 K \cite{McWilliams2016}.
\label{fig:BG_Liq}}
}\end{figure}

Figure \ref{fig:BG_Liq} shows the estimates of fundamental gap, computed according to Eq. \ref{eq:gapApp}, for different isotherms of liquid hydrogen. The gap gradually decreases with pressure and depends on both temperature and density as can be seen in the inset. Below the critical temperature of the LLPT, the gap closure coincides with the beginning of the coexistence region, as indicated by colored rectangles. In this region the accuracy of the estimated gap is uncertain, since during the simulation at constant volume the system dynamically switches from atomic to molecular states and back.  Note that at all temperatures the gap decreases linearly with pressure, with the slope becoming steeper as temperature increases. 

\begin{figure}
\center
\begin{minipage}[b]{\columnwidth}
\subcaption{}
\includegraphics[width=\columnwidth]{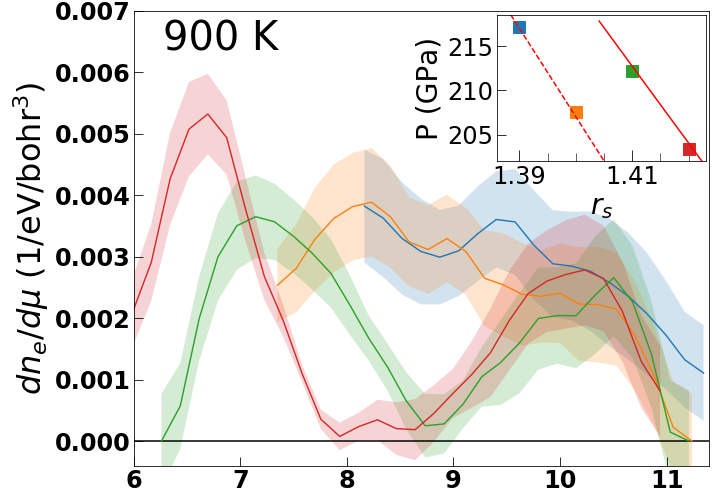}
\end{minipage}
\begin{minipage}[b]{\columnwidth}
\subcaption{}
\includegraphics[width=\columnwidth]{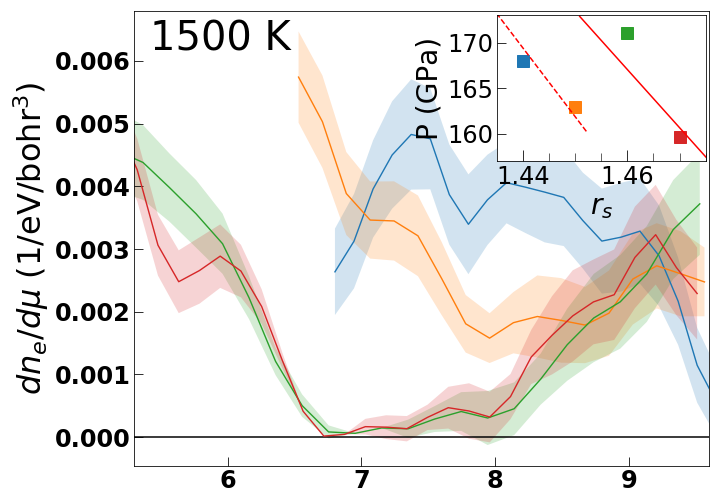}
\end{minipage}
\begin{minipage}[b]{\columnwidth}
\subcaption{}
\includegraphics[width=\columnwidth]{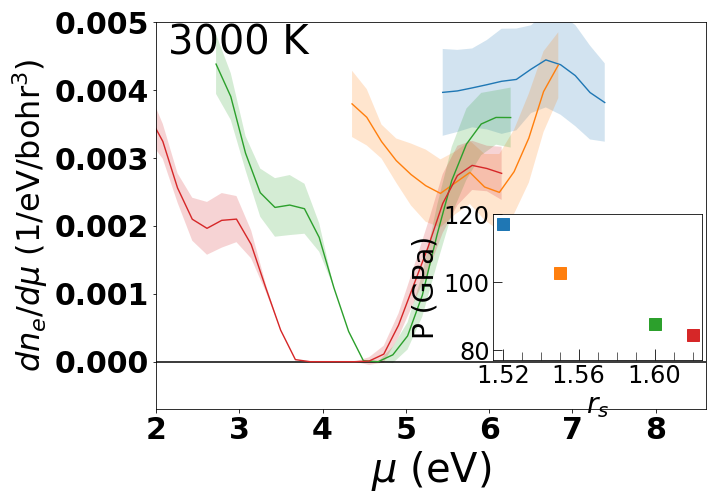}
\end{minipage}
\caption{\small{Density of states of liquid hydrogen near the band edge at densities near the gap closure for three isotherms: (a) 900 K, (b) 1500 K and (c) 3000 K. The insets show the equation of state as reported in \cite{Pierleoni2017}. The dashed and solid red lines indicate the atomic and molecular region respectively. The colors of the DOS match the colors of points in the insets. 
\label{fig:DOS_QMC}}
}\end{figure}

From the electronic density of states (DOS), shown in Fig. \ref{fig:DOS_QMC}, 
%near the valence and conduction band edges, obtained  within GCTABC-QMC, 
we obtain important information on the character of the transition.
%Figure \ref{fig:DOS_QMC} shows the DOS at three isotherms 
%for densities around the gap closure. 
Below the critical temperature, at 1500 K and 900 K, we show the DOS at four densities around the LLPT. The equation of state is plotted on the inset as reported in Ref. \cite{Pierleoni2017}. 
On the molecular side (higher $r_s$), the DOS has a clearly
visible gap where the density of states is almost vanishing.
Although the finite system size and the
finite number of nuclear configurations underlying our calculations
do not allow us to
distinguish between a strictly vanishing DOS 
in the thermodynamic limit or a semiconductor DOS containing a small fraction of
disorder-like (impurity) states inside the gap, the shape of the DOS changes abruptly
from the molecular to the atomic liquid. 
The width of the gap 
continuously follows the
molecular branch inside the coexistence region,
strongly supporting
a scenario where the gap vanishes discontinuously at the molecular-atomic transition.

Above the critical temperature, at 3000 K, the DOS reflects the mixed  molecular/atomic character
of the liquid. Although 
more calculations between $r_s=1.6$ (green point), where the gap is 0.8 eV,
and $r_s=1.55$ (orange point), where the gap is closed, are needed to precisely locate the
closure of the gap, the strong correlation of the DOS with the molecular character
suggests that the gap closes continuously as a function of density (see inset
of Fig. \ref{fig:BG_Liq}), and, thus, also as a function of pressure.

The molecular fraction as a function of pressure has been analysed in Ref. \cite{Pierleoni2017} 
based on different criteria. 
Although all estimators used in that reference
show the onset of molecular dissociation within the gapped liquid, the
values of the molecular fraction are sensitive to the estimator. This implies that,
we cannot determine whether a gapped atomic liquid is reached before gap closure.

\begin{figure*}[t]
\begin{minipage}[b]{\columnwidth}
\subcaption{}
\includegraphics[width=\columnwidth]{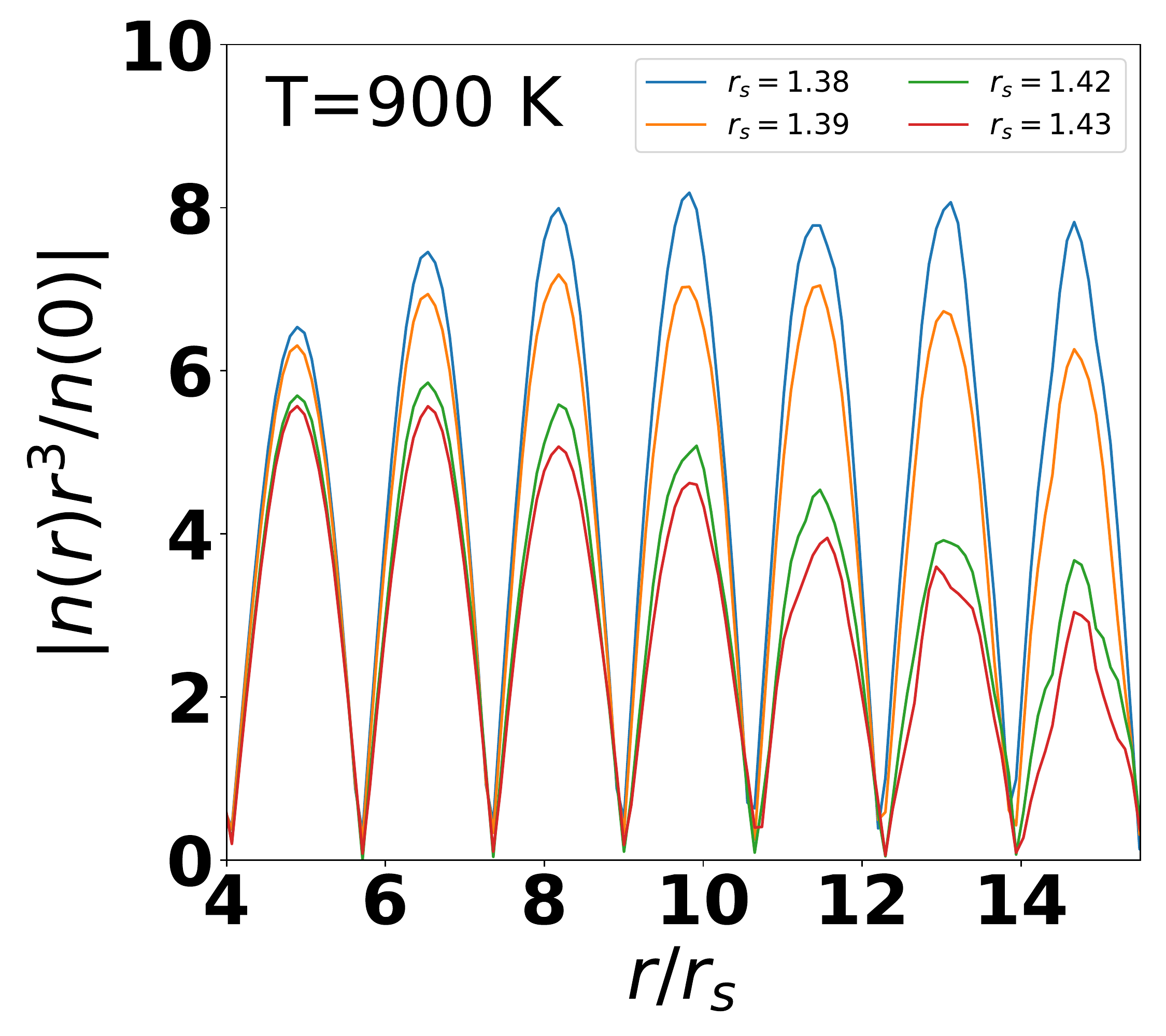}
\end{minipage}
\begin{minipage}[b]{\columnwidth}
\subcaption{}
\includegraphics[width=\columnwidth]{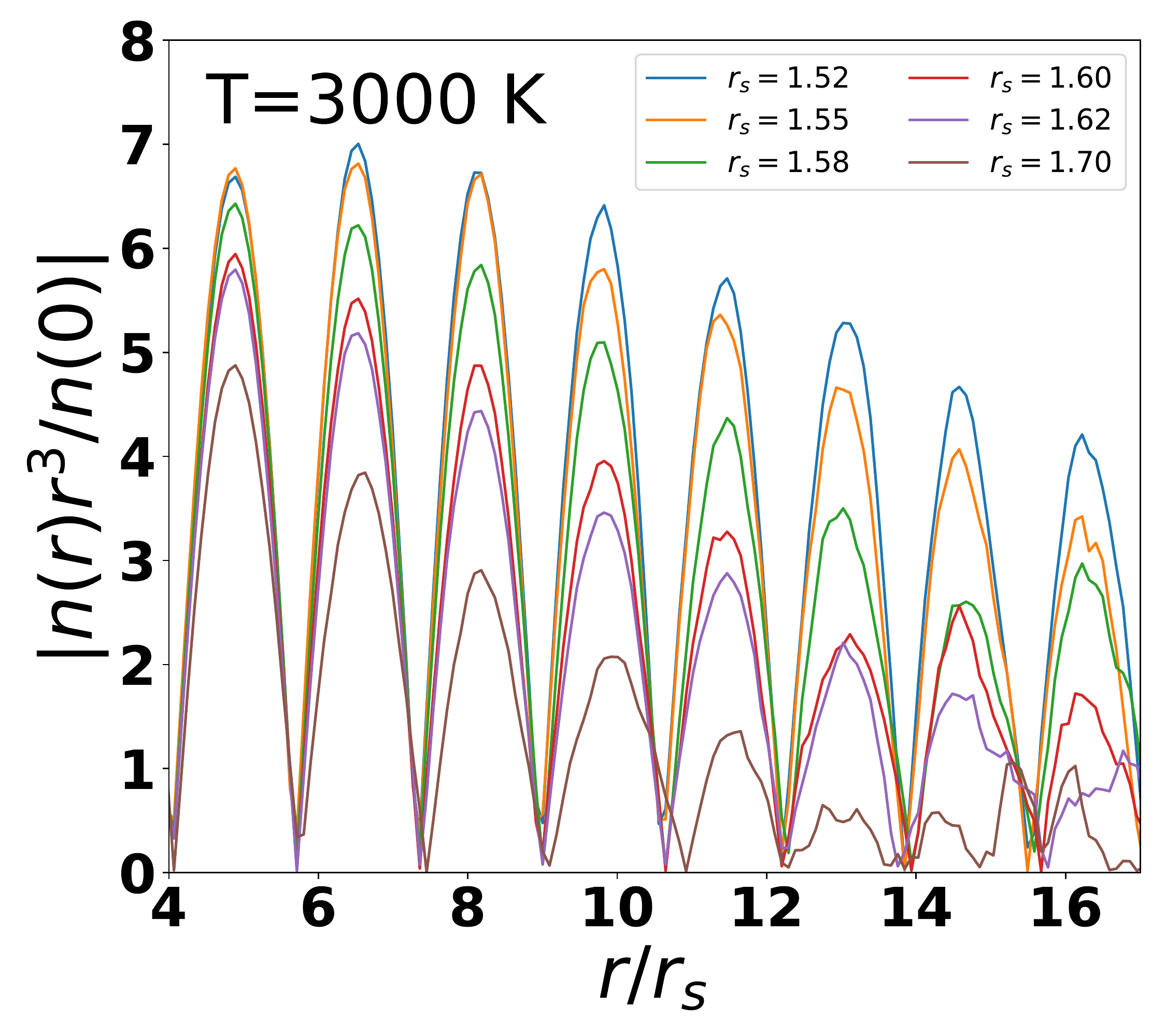}
\end{minipage}
\caption{\small{The absolute value of the off-diagonal part of the reduced single particle density matrix, $n(r)$,
\cite{Pierleoni2018} multiplied by $r^3$ as a function of distance $r$ for various densities around
gap closure. (a) At $T=900$,  when the gap vanishes for $r_s \lesssim 1.42$, as the liquid crosses the LLPT, the $n(r)$ changes indicating more delocalized, Fermi-liquid like behaviour.  (b) At $T=3000$, since $n(r)$ decays faster than $r^{-3}$ at all densities, the electron liquid remains
localized even when the gap vanishes ($r_s \lesssim 1.55$).  }
\label{fig:nr_t3kk}}
\end{figure*}

In a normal solid, the vanishing of the gap usually implies a IM transition, e.g.
a transition from a state of almost vanishing conductivity to a state where
electronic conductivity is only limited by nuclear (phononic) motion and/or impurities.
However, a liquid is similar to a disordered system; the vanishing of the gap does not necessarily
imply the existence of extended states at the Fermi level needed for transport.
Further information on the extended/localized character of the states around the Fermi level
is needed, in order to determine the insulating or metallic character of the liquid
\cite{KohnInsulat}.

The QMC results for electronic momentum distribution
and its Fourier transform, the reduced single particle density matrix
along the LLPT have been presented and analysed in Ref. \cite{Pierleoni2018}.
The asymptotic behavior
of the off-diagonal part of the single particle density matrix,
$n(r)$, at large distances $r$ discriminates between extended and localized states, the latter decaying
faster than $r^{-3}$ \cite{Pierleoni2016}.
Below the critical temperature,
the off-diagonal part of the density matrix abruptly changes from a roughly exponentially decay
in the molecular phase to an algebraic Fermi-liquid like behavior in the atomic liquid as seen in Fig. \ref{fig:nr_t3kk}(a).
The closure of the gap induces a IM transition which occurs together 
with the thermodynamic molecular-atomic transition.

Above the critical point the situation is different,
the momentum distribution changes smoothly with density as seen in Fig. \ref{fig:nr_t3kk}(b).
We see that $n(r)$ decays faster than $r^{-3}$ at $T=3000$K for
the densities below and above gap closure,
implying a localized electron liquid. At gap closure, the liquid enters a gapless 
localized phase.
This enables absorption at arbitrary low energies.
We expect no sharp IM transition but a cross-over to the metallic liquid, 
since delocalization increases smoothly with density or pressure.
Indeed, conductivity as well as other transport properties obtained within DFT
change smoothly \cite{Morales2010,Rillo2019}.
We further note, that the DOS after gap closure shown in Fig. \ref{fig:DOS_QMC} actually resembles that of a dirty
semiconductor containing localized (disordered) states inside the gap.

We now compare our results to experimental estimates  \cite{Celliers2018,Nellis1999,McWilliams2016}. Cellier et al. \cite{Celliers2018} have extracted the gap based on the empirical relations to the refractive index data (as discussed in details in SM of Ref. \cite{Celliers2018}). The agreement with our results is quite good, although the experiment is for deuterium, however our result does not support the extrapolation procedure provided in the paper. Another estimate of the gap is based on the semiconductor model of thermally activated conductivity, $\sigma$, \cite{Nellis1999} 
\beq
\sigma(\rho,T)=\sigma_0 \exp(-E_g(\rho)/2k_BT),
\eeq
where $\sigma_0$ is the limiting value of conductivity and $E_g(\rho)$ is the energy gap, assumed to depend linearly on the density $\rho$ and independent of the temperature $T$. % $k_B$ is the Boltzmann constant. 
Note that in the original paper of Nellis et al. \cite{Nellis1999} the choice of the limiting value of conductivity, $\sigma_0$ was arbitrary, $\sigma_0$ is a free parameter that varied between 66 to 300 ($\Omega$ cm)$^{-1}$, a value typical of liquid semiconductors \cite{MOTT1972}. In Fig. \ref{fig:BG_Liq} we report results of Nellis et al. \cite{Nellis1999} reanalysed by Knudson et al. \cite{Knudson2018}  who used a different equation of state \cite{Kerley032003} and different $\sigma_0$. They assumed that hydrogen before the transition behaves like a fluid semiconductor, where the conductivity is progressively increased upon the closure of the gap with density. The value of $\sigma_0$ was chosen so the resulting gap was not negative. The gap is assumed to weakly depend on temperature which was not measured and, according to the latest equation of state \cite{Kerley032003}, varied between 2000-3000 K, increasing towards the higher pressure \cite{Knudson2018}. Below the critical temperature, our results do not fully support this model, as the QMC density of states increases rapidly at the transition (see Fig. \ref{fig:DOS_QMC})  and our gap is temperature dependent. Above the critical temperature, we do not have enough data to assess the model, as we would need at least three isotherms, but the form of the DOS discussed
above supports the use of a semiconductor model.   
%\CP{We can mention that HSE conductivity at 3000K has been shown to be larger than Nellis's results in ref[40] (see SM figure 3). Correcting for the difference between HSE and QMC gap will reduce the prediction probably bringing the two sets closer. Do we have values of sigma with scissor corrected gap?} \vitaly{we would need to recompute the conductivity at 3000K as we don't have the converged HSE gap}

\begin{figure}[t]
%\center
%\begin{minipage}[b]{0.55\columnwidth}
\includegraphics[width=\columnwidth]{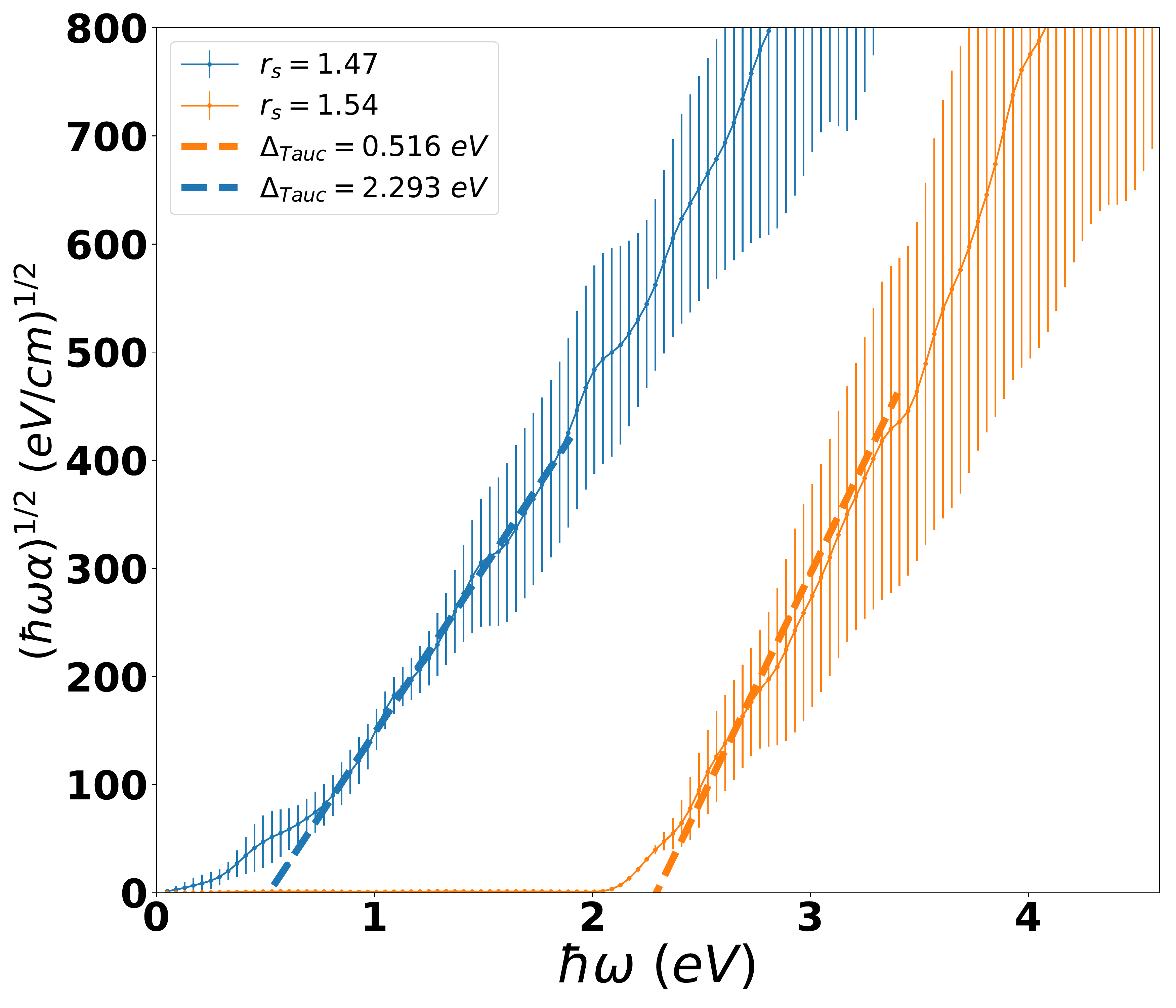}
%\end{minipage}
\caption{\small{Tauc analysis of the absorption profiles, computed with DFT-HSE for liquid hydrogen at $T=1500$ K and two densities: $r_s=1.54$ and 1.47
\label{fig:TaucT1500}}
}\end{figure}

Lastly, analysing the absorption profile with the Tauc model \cite{Tauc1968}, McWilliams et al. \cite{McWilliams2016} have reported the gap value of 0.9 eV at 2400 K and 140 GPa \cite{McWilliams2016}. To assess the validity of this model, we analysed several theoretical absorption profiles with DFT-HSE for two densities ($r_s=1.54$ and 1.47) at $T=1500$ K. We found that the fitting of the theoretical absorption to the Tauc model slightly overestimates the values of the gaps (by $\sim 0.3$ eV), computed at the same level of approximation as optical properties, e.g. DFT - HSE. This is shown in figure \ref{fig:TaucT1500}. However, the Tauc model gives good agreement with the QMC gap, indicating the possibility of error cancellation, when calculating the spectra.  

\begin{figure*}[t]
\center
\begin{minipage}[b]{\columnwidth}
\subcaption{}
\includegraphics[width=\columnwidth]{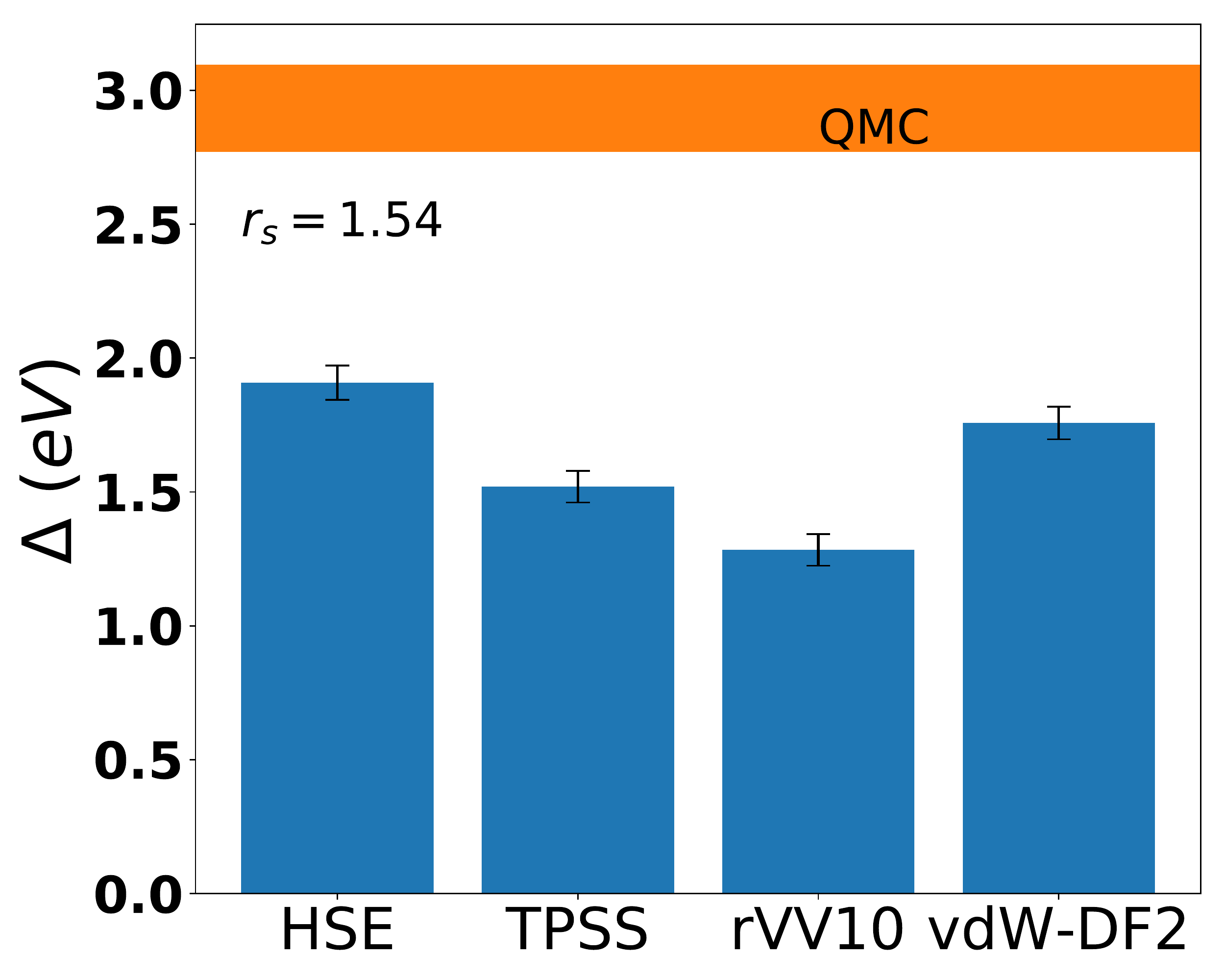}
\end{minipage}
\begin{minipage}[b]{\columnwidth}
\subcaption{}
\includegraphics[width=\columnwidth]{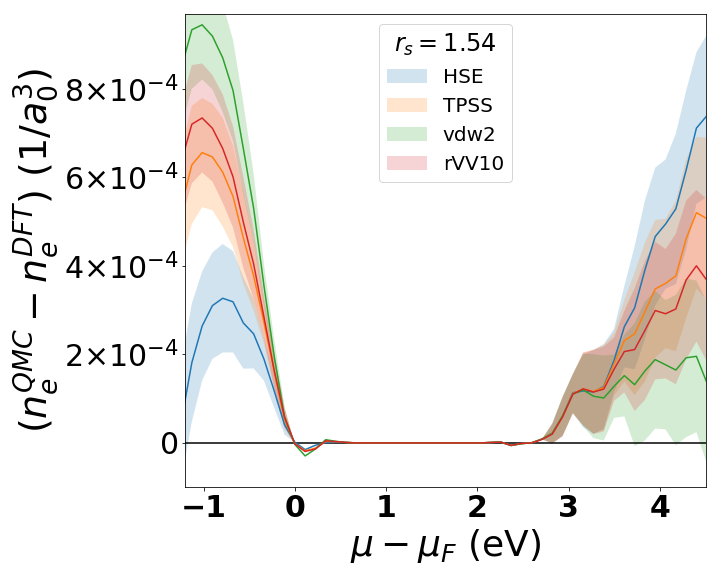}
\end{minipage}
\caption{\small{ (a) The fundamental gap calculated with different DFT exchange correlation functionals compared with the QMC gap at $T=1500$ $K$ and $r_s=1.54$. The orange horizontal bar is the RQMC-GCTABC thermal gap with its statistical uncertainty given by its width. (b) Difference between the integrated density of states of QMC and DFT. A ``scissor correction'' on the horizontal axis from the gap value has been applied to the DFT profiles before subtracting them from the QMC profile, $\mu_F$ has been set by the maximum of the valence band.
\label{fig:DOS_DFT_QMC}}
}\end{figure*}

\subsection{Benchmark of XC approximations}

\begin{figure*}[t]
\center
\begin{minipage}[b]{\columnwidth}
\subcaption{}
\includegraphics[width=\columnwidth]{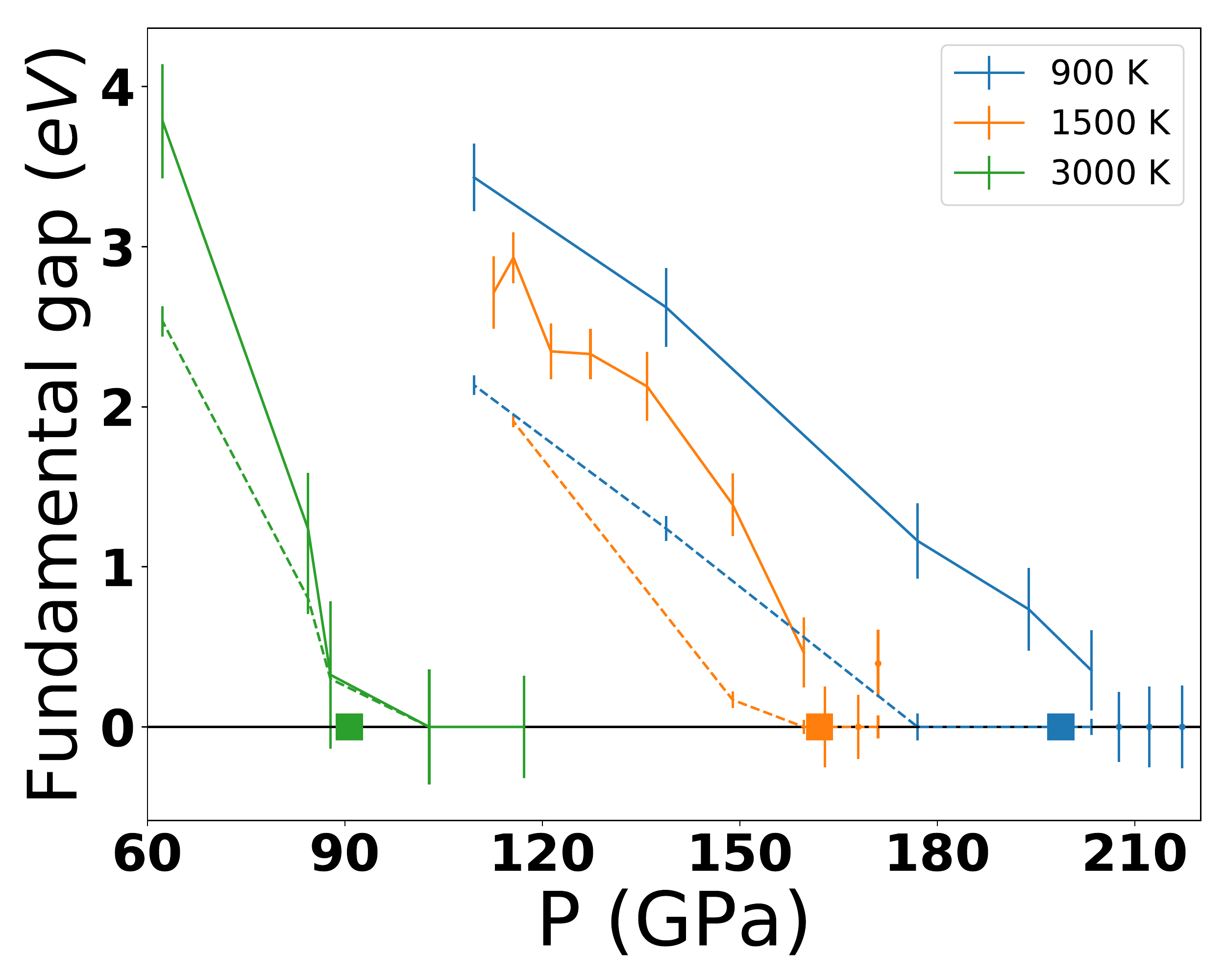}
\end{minipage}
\begin{minipage}[b]{\columnwidth}
\subcaption{}
\includegraphics[width=\columnwidth]{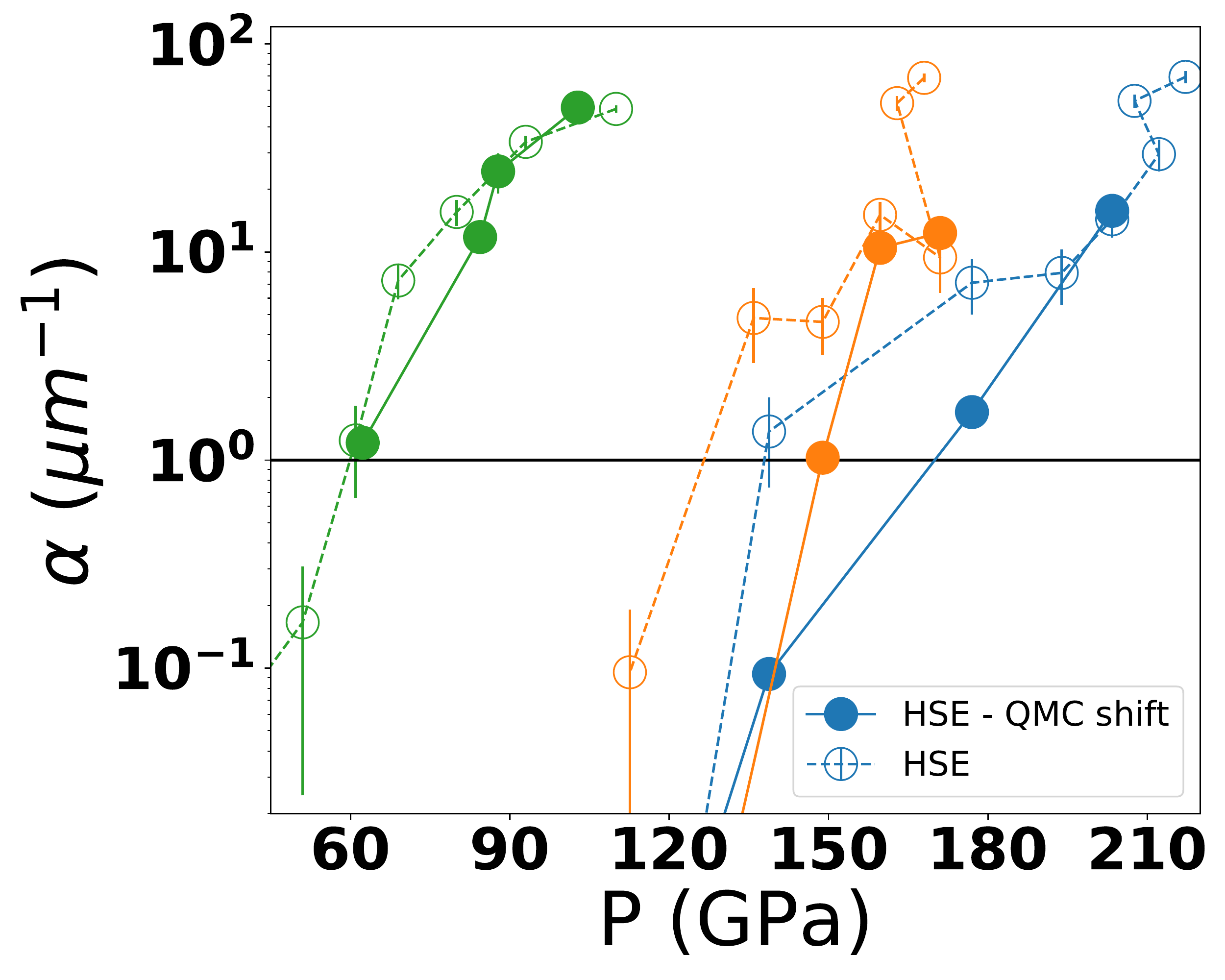}
\end{minipage}
\caption{\small{(a) The HSE and QMC band gaps along three isotherms. The dashed lines are the HSE values and solid lines are the QMC results. Squares indicate pressures at which the reflectivity is 0.3 according to Ref. \cite{Rillo2019}. (b) Absorption at $\omega = 2.3$  eV along the $T=1500$ K isotherm. The dashed lines are the HSE values reported in Ref. \cite{Rillo2019} and the solid lines are computed using the QMC corrected band gaps. 
\label{fig:Abs_R_Shift}}
}\end{figure*}

Figure \ref{fig:DOS_DFT_QMC}(a) shows the value of the gap using different DFT functionals compared to the thermal gap from RQMC-GCTABS at 1500 K and $r_s=1.54$. Four functionals were considered: non-local and semi-local Van-der-Waals density functionals rVV10 \cite{rVV102013} and vdW-DF2 \cite{Lee2010}, semi-local meta-GGA TPSS functional \cite{TPSS2003}, and non-local hybrid HSE \cite{Heyd2005}. The differences are on the order of $\sim 1$ eV with HSE and vdW-DF2 being the closest to the QMC prediction, a result similar to solid hydrogen \cite{Gorelov2020}. We also access the accuracy of the intensity of optical properties computed with different XC functionals. With QMC we do not have a direct access to the optical properties, but to a large extent they are defined by the density of states. In Figure \ref{fig:DOS_DFT_QMC}(b) we show for each DFT approximation the difference of the integrated density of states between QMC and gap-corrected DFT DOS (we correct the value of the gap to match the QMC one).  As with the gap comparison, HSE and vdW-DF2 perform better than the others. Note that the vdW-DF2 performs better on the conduction side and HSE is better on the valence side. Therefore, considering the computational cost of non-local hybrid functionals, it might be advantageous to use semi-local vdW-DF2. Another important conclusion, coming from Fig. \ref{fig:DOS_DFT_QMC} is that correcting just the gap error in DFT does not guarantee that the intensities of the spectra are accurate, they are probably underestimated within the  XC approximations since the difference between the QMC and DFT DOS is always positive, implying that there are fewer states contributing to the DFT spectrum.

\subsection{Optical properties}

%\CP{You do not discuss at all whether you use semiclassical averaging or QA. I think we should mention this.}

Comparing the HSE and QMC gaps versus pressure along the three isotherms in Figure \ref{fig:Abs_R_Shift}(a), we notice a constant shift of roughly $\sim 1$ eV between HSE and QMC gaps below the critical temperature. At the highest temperature, $T=3000$ K, the difference decreases with increased pressure. The gap closes at the same values of pressure with DFT and QMC at 3000K and 15000K, while at $T=900$ K the HSE gap closes at 180GPa while the QMC gap closes at $\sim 200GPa$. %Such a difference in the gap will result in a shift \CP{do you mean an horizontal or vertical shift? Probably we should replace shift with change} in the theoretical absorption profile reported previously \cite{Rillo2019}. 
Figure \ref{fig:Abs_R_Shift}(b) shows that when shifting the HSE eigenvalues to match the QMC gap, the value of absorption at 2.3 eV plotted as a function of pressure decreases, with the shift being more pronounced at lower pressures and lower temperatures. At high pressure, when the gap is already closed, we cannot apply the "scissor correction" and the value of absorption at 2.3 eV will be based purely on the optical transition intensity, which we cannot currently calculate within QMC. The DOS in Fig. \ref{fig:DOS_DFT_QMC}(b) suggests that the DFT optical intensities might be underestimated, as the QMC DOS is higher near the gap, therefore we expect an error cancellation between an underestimated band gap and underestimated intensities to occur. 

The reflectivity can be reanalysed in the same manner. Consistent with decreased absorption at lower pressure, reflectivity decreases as well. However, we do not provide the same analyses here for the following reason: we are interested at the IM transition, which characterised by the typical values of reflectivity $\sim 0.3$, the gap at this value is either small or already closed and the correction will be negligible. The pressure at which reflectivity reaches 0.3, according to \cite{Rillo2019} is reported on Fig. \ref{fig:Abs_R_Shift}(a) as colored squares. Therefore, correcting the gap for the reflectivity will not produce significant changes; the accuracy of optical properties will be determined by the accuracy of the intensities of optical transitions.  %\vitaly{say that he gap at 0.3 is already closed} \CP{I think you already explained me this but I forgot and I find the explanation here unclear.}
%This fact will cancel the band gap shift, bringing the absorption closer to DFT-HSE values, indicated by dashed line on fig. \ref{fig:Abs_R_Shift}(b).

The pressure at which hydrogen turns opaque was attributed in several experiments to correspond to the absorption of $\sim1$ $(\mu m)^{-1}$ \cite{Celliers2018,Knudson2015,McWilliams2016}. Based on our QMC shifted HSE absorption, we predict that the onset of absorption will shift to higher pressures, with respect to the previously reported ones \cite{Rillo2019}. This again indicates that the absorption intensities might be underestimated within the HSE XC approximation.% \vitaly{which again indicates an error cancellation between an underestimated band gap and underestimated intensities.} 

%\david{Not sure I understand this paragraph. Needs clarification.} \CP{I agree} On the other hand, closure of the gap, being an indicator of IM transition, perfectly underlines importance of the experimentally observed increase of reflectivity and coincides with the previously reported \cite{Pierleoni2016} structural phase transition. Hence the liquid-liquid phase transition is also an IM transition accompanied by the increase of the reflectivity and conductivity.

%\begin{figure*}[t]
%\center
%\begin{minipage}[b]{\columnwidth}
%\includegraphics[width=1.7\columnwidth]{Figures/H2Liq_PT.pdf}
%\end{minipage}
%\caption{\small{Reanalysed phase diagram with shifted onset of absorption line (blue dashed line with open circles). Open points indicate the experimental onset of absorption or temperature plateau. Solid points indicates the P-T conditions of the reflective sample observed experimentally. Light blue solid points connected by the line indicate the closing of the QMC gap. The dashed line is the liquid-liquid phase transition reported at Ref. \cite{Pierleoni2016}.
%\label{fig:BG_PT}}
%}\end{figure*}

\section{Conclusion}\label{sec:conclusions}

In this paper, we have reported values of the fundamental gap across the pressure induced molecular dissociation region in hydrogen using a newly developed QMC method \cite{Yang2020}. 
The main finding is that gap closure 
strongly correlates with the beginning of the molecular dissociation transition. 
Below the critical temperature, the gap closure occurs abruptly, with a small discontinuity
reflecting the weak first-order themodynamic transition.
Above the critical temperature, molecular dissociation begins before the closing of the gap.
Despite the liquid becoming gapless, the change from insulating to metallic behavior
occurs progressively. %\CP{Is this last sentence applying to both below and above Tc?} 
On the basis of our QMC density of states, we have further
benchmarked different DFT functionals and found that all considered functionals underestimate the gap. After applying a scissor correction on the energy spectrum, HSE XC optical transition intensities, previously found to agree with experiments \cite{Rillo2019}, are now lower and in less good agreement with experiments (see Fig. \ref{fig:Abs_R_Shift}(b)).
Our analysis of the DOS at the band edges (see Fig. \ref{fig:DOS_DFT_QMC}(b)) suggests that the QMC spectrum have more states than DFT ones hence should have larger intensity, possibly restoring the agreement with experiments. In other words, our analysis suggests that the previously observed agreement between HSE optical profiles and experiments \cite{Rillo2019} profited by error cancellation. This observation remains to be established by a more systematic investigation.  %\vitaly{As a result, optical intensities and the onset in absorption profiles might be underestimated based on HSE XC approximation, which indicates an error cancellation within the DFT optical properties and explains the agreement with experiment.}

\begin{acknowledgments}

D.M.C. was supported by DOE Grant NA DE-NA0001789 and by the Fondation NanoSciences (Grenoble). V.G. and C.P. were supported by the Agence Nationale de la Recherche (ANR) France, under the program ``Accueil de Chercheurs de Haut Niveau 2015'' project: HyLightExtreme. Computer time was provided by the PRACE Project 2016143296, ISCRAB (IsB17\_MMCRHY) computer allocation at CINECA Italy, the high-performance computer resources from Grand Equipement National de Calcul Intensif (GENCI) Allocation 2018-A0030910282, and by
the Froggy platform of CIMENT, Grenoble (Rh{\^o}ne-Alpes CPER07-13 CIRA and ANR-10-EQPX-29-01).
\end{acknowledgments}

\bibliographystyle{apsrev4-1}
\bibliography{main}
\end{document}